\documentclass[fleqn,usenatbib]{mnras}
\usepackage[T1]{fontenc}
\usepackage{ae,aecompl}

\usepackage{graphicx}	
\usepackage{amsmath}	
\usepackage{amssymb}	
\usepackage{xcolor} 
\usepackage{cuted}
\usepackage{times}

\newcommand{\msun}{{\rm M}_\odot}

\newcommand{\feh}{\mathrm{[Fe/H]}}
\newcommand{\epz}{\epsilon_\mathrm{z}}

\newcommand{\empempall}{$M_{\mathrm{EMP, comp}}$-to-$M_\mathrm{EMP} (<300\mathrm{kpc})$\,}
\newcommand{\empall}{$M_{\mathrm{EMP, comp}}$-to-$M_{\mathrm{tot, comp}}$\,}

\newcommand{\empallr}{$M_{\mathrm{EMP}}(r)$-to-$M_{\mathrm{tot}}(r)$\,}

\title[Location of EMPs in TNG50 MW/M31 analogues]{Where are the extremely metal-poor stars in the Milky Way and Andromeda? Expectations from TNG50}
\author[L.-H. Chen et al.]{
Li-Hsin Chen$^{1,2}$, Annalisa Pillepich$^{3}$, Simon C. O. Glover$^{1}$, and Ralf S. Klessen$^{1,4}$
\\
$^{1}$Universit\"{a}t Heidelberg, Zentrum f\"{u}r Astronomie, Institut f\"{u}r Theoretische Astrophysik, Albert-Ueberle-Str.\ 2, \\ 69120 Heidelberg, Germany\\
$^{2}$International Max Planck Research School for Astronomy and Cosmic Physics at the University of Heidelberg (IMPRS-HD), \\ K\"{o}nigstuhl 17, D-69117 Heidelberg, Germany \\
$^{3}$Max-Planck-Institut f\"{u}r Astronomie, K\"{o}nigstuhl 17, D-69117 Heidelberg, Germany \\
$^{4}$Universit\"{a}t Heidelbergt Heidelberg, Interdisziplin\"{a}res Zentrum f\"{u}r Wissenschaftliches Rechnen, Im Neuenheimer Feld 225, \\ D-69120 Heidelberg, Germany\\
}

\date{}

\begin{document}
\label{firstpage}
\pagerange{\pageref{firstpage}--\pageref{lastpage}}
\maketitle

\begin{abstract}
We analyse the location of extremely metal-poor stars (EMPs, $\feh < -3$) in 198 Milky Way (MW)/M31-like galaxies at $z=0$ in the TNG50 simulation. Each system is divided into four kinematically-defined morphological stellar components based on stellar circularity and galactocentric distance, namely bulge, cold disk, warm disk, and stellar halo, in addition to satellites (with stellar mass $\ge5\times10^6\,\msun$). According to TNG50 and across all simulated systems, the stellar halo of the main galaxy and satellites present the highest frequency of EMPs (largest \empall stellar mass ratio), and thus the highest chances of finding them. Such frequency is larger in lower-mass than high-mass satellites. Moreover, TNG50 predicts that the stellar halo of the main galaxy always hosts and thus contributes the majority of the EMPs of the system. Namely, it has the highest mass ratio  of EMPs in it to all the EMPs in the system (largest \empempall). However, notably, we also find that 33 MW/M31-like galaxies in TNG50 have cold disks that contribute more than 10 per cent to the total EMP mass, each with $\gtrsim 10^{6.5-7}\, \msun$ of EMPs in cold circular orbits. These qualitative statements do not depend on the precise definition of EMP stars, i.e. on  the adopted metallicity threshold. The results of this work provide a theoretical prediction for the location of EMP stars from both a spatial and kinematic perspective and across an unprecedented number of well-resolved MW/M31-like systems.
\end{abstract}


\section{Introduction}
\label{sec:emp_intro}
Extremely metal-poor ($\feh < -3$\footnote{We adopt the notation of $\feh = \mathrm{log}_{10} (N_\mathrm{Fe}/N_\mathrm{H}) - \mathrm{log}_{10} (N_\mathrm{Fe,\odot}/N_\mathrm{H,\odot})$, where $N_\mathrm{Fe}$ and $N_\mathrm{H}$ are the fractional abundances of iron and hydrogen, respectively, and $N_\mathrm{Fe,\odot}$ and $N_\mathrm{H,\odot}$ correspond to the solar abundances \citep{Asplund:2009aa}.}) stars \citep[EMPs][]{Beers:2005aa} are one of the best candidates to study the first generation, metal-free stars (Population~III or Pop~III stars). Due to their low metallicity, it is likely that the gas where these EMP stars formed was polluted by only a handful of or even one only Pop~III star \citep{Ishigaki:2014aa,Ishigaki:2018aa,Keller:2014aa,Tominaga:2014aa,Ji:2015aa,Placco:2015aa,Placco:2016aa,Fraser:2017aa,Magg:2018aa,Hartwig:2018aa}.

Numerical simulations of stars have shown that a wide range of stellar metallicities can in fact be realised by second generation stars. \citet{Yoshida:2004aa} found that gas metallicity can reach values as high as $10^{-4}Z_\odot$ by pair instability supernovae (SNe) at $z \sim 15-20$ already. \citet{SmithB:2007aa} performed a series of hydrodynamical simulations in a 300 $h^{-1}$kpc box with different fixed gas metallicities ($Z = [0, 10^{-4}, 10^{-3}, 10^{-2}]Z_\odot$) and found that fragmentation started to occur at $10^{-3}Z_\odot$, which indicates a transition to the formation of low-mass stars.
\citet{Greif:2010aa} carried out a set of hydrodynamical simulations to study the collapse of minihaloes at $z \sim 10$ after the gas in them had been enriched to $\feh \sim -3$ by Pop~III SNe.
More recently, \citet{Jeon:2014ab} showed that a single core-collapse Pop~III SN can enrich the gas to $\feh \sim -4$. \citet{SmithB:2015aa} simulated two Pop~III SNe in a box of 500 $h^{-1}$kpc and found that more than 10 minihaloes are enriched by the SNe and have metallicities from $10^{-4}Z_\odot$ to $10^{-2}Z_\odot$. \citet{Chiaki:2020aa} studied the formation of carbon-enhanced metal-poor star in a halo that collpases at $z=10$ whereby extreme cases of stars with $\feh \sim -9.25$ were reported. \citet{Magg:2022aa} modelled the formation of the first stars and their deaths as pair instability supernovae to study metal mixing in mini haloes. They found that most of the second generation stars have $\feh \leq -3$ with a few that have $\feh \geq -2$.

In the past few decades, many researchers have devoted energy and time to the search for EMPs in the system of our own Galaxy \citep[the Milky Way (MW):][]{Beers:2005aa,Lai:2008aa,Norris:2013aa,Frebel:2015aa,Ji:2016aa,HansenCJ:2020aa}.
For example, \citet{Yong:2013aa} reported 190 metal-poor stars, where 10 of them have $\feh \leq -3.5$.
\citet{Roederer:2014aa} analysed 313 metal-poor stars including 19 stars with $\feh \leq -3.5$.
Rare stars with $\feh \leq -6$ were also discovered in recent years  \citep{Keller:2014aa,Frebel:2019aa,Nordlander:2019aa}. 

On the other hand, large stellar surveys have helped to build a statistical census of the stars in the MW and their metallicity distribution. Notably, \citet{Hayden:2014aa} analysed 20,000 stars and derived the mean metallicity map of the Galaxy from the stellar spectroscopic APOGEE survey, which covers the stellar disk and the stellar halo out to a few kpc in height from the mid plane, and observed stars down to metallicities of $\gtrsim 0.01 Z_\odot$.
\citet{Chiti:2021aa,Chiti:2021ab} studied the metallicity distribution function (MDF) of the MW and presented metallicities of $\sim 28,000$ stars down to $\feh \lesssim -3.75$ from the SkyMapper Southern Survey. 
The Pristine survey \citep{Starkenburg:2017ab}, where the Ca H\&K lines are used to infer the photometric metallicity with good accuracy down to the extremely metal-poor regime of $\feh < -3.0$, covers $\sim 1000$ deg$^2$ of the sky. With follow-up spectroscopic studies by \citealt{Youakim:2017aa} and \citealt{Lardo:2021aa}, about $1000-1200$ EMP stars have been identified and a metallicity floor of $\feh = -4.66$ \citep{Starkenburg:2018aa} has been reached.



In terms of their spatial or orbital location, EMPs are being identified almost everywhere in our Galaxy. The Galactic bulge is considered to be a potential site of possible Pop~III survivors \citep{Schlaufman:2014aa} and many very metal-poor ($\feh < -2$) stars have been discovered there \citep{Howes:2015aa,GarciaPerez:2016aa,lamb:2017aa,Arentsen:2020aa,Arentsen:2021aa}. Yet, \citet{Lucey:2021aa} conducted an orbital analysis of stars that were identified -- spatially -- in the Galactic bulge and found that half of their sampled stars in fact have $>50\%$ possibility of being halo stars that happen to be crossing the Galactic bulge. The fraction of such halo stars increases with decreasing metallicity (in the range of $-3 < \feh < 0.5$). \citet{Kielty:2021aa} presented high-resolution spectra of 30 metal-poor stars and most of them are in the Galactic halo. More generally, \citet{Venn:2020aa} performed spectroscopic analysis of metal-poor stars and found various types of orbital features, with the exciting and somewhat unexpected finding of EMPs with near-circular orbits \citep{Caffau:2012aa,Schlaufman:2018aa,DiMatteo:2020aa,Sestito:2020aa,Mardini:2022aa}. 

From a theoretical perspective, the common picture is that the majority of the lowest-metallicity stars in disk galaxies should be distributed as an isotropic, pressure-supported component, i.e. should be located within the bulges and stellar haloes. This is because these stars are thought to have been either accreted from the early building blocks of the assembling object (as Pop~III, see above, or subsequent Pop~II stars) or to have been later brought in by faint, low-mass dwarf galaxies \citep{Brook:2007aa,Gao:2010aa,Salvadori:2007aa,Salvadori:2010aa,deBennassuti:2014aa,Hartwig:2015aa}. 
In fact, quantitative predictions are scarce. 
\citet{Scannapieco:2006aa} found that the mass fraction of Pop~III stars in the simulated Galactic halo is comparable to it in the bulge from a study which combined a N-body simulation and a semi-analytic model. 
\citet{Tumlinson:2010aa} has shown that the dark matter (DM) particle is associated with star particles much earlier than the DM particle becomes associated with the host halo (see their Figure 3 and the corresponding text). The time difference between these two events can be as large as $\Delta z = 10$. This time difference reflects that the build-up of galaxies is hierarchical and stars we observe today may have been accreted.
\citet{Komiya:2016aa} studied the Pop~III survivors that have escaped from the mini-haloes where they form could locate at the outskirts of the MW ($\sim 100\,$kpc from the Galactic centre).
With cosmological hydrodynamical simulations of MW-like galaxies, \citet{Starkenburg:2017aa} analysed stars with $\feh < -2.5$ in Local Group analogues in the APOSTLE simulations \citep{Sawala:2016aa, Fattahi:2016aa} 
and found that $\sim 60 \%$ of these metal-poor stars reside at $>8\,$kpc from the galactic centres. However, this quantitative result is dependent on the underlying metal-mixing model. 
\citet{El-Badry:2018af} showed that the median formation redshift of stars with $\feh < -2$ is $z=5$ from zoom-in simulations of three FIRE MW-mass galaxies \citep{Hopkins:2014aa}.
Upon analysing the stellar orbits of low-metallicity ($\feh \leq -2.5 $) stars in MW-like galaxies taken from the NIHAO-UHD project \citep{Buck:2020aa}, \citet{Sestito:2021aa} found that low-metallicity stars that are accreted into the MW progenitor halo in the first few Gyr can populate either prograde or retrograde orbits; on the other hand, low-metallicity stars that merged when the disk in the MW progenitor formed occupy mostly the prograde orbits, all this depending on the mass ratio of the past mergers. 
Recently, \cite{Pakmor:2022aa} studied the location and rates of star formation for stars below $0.1 Z_\odot$ in the large-volume cosmological simulation TNG50 \citep{Pillepich:2019aa, Nelson:2019ab} across the galaxy mass spectrum: however, they did not extend their study below $0.01 Z_\odot$. 

In this work, we build upon these recent results and use the TNG50 cosmological simulation, the highest-resolution run of the IllustrisTNG project\footnote{\url{www.tng-project.org}}, to provide quantitative expectations as to where, and how easily, EMPs may be found in our Galaxy and in Andromeda. TNG50 returns about 200 galaxies that can be considered analogues of the MW and M31 at $z=0$ (Pillepich et al. in preparation), i.e. an unprecedented simulated sample with a stellar particle mass resolution of $8.5\times 10^4\, \msun$. We can thus provide predictions for the frequency of EMPs within the various galactic morphological components (bulge, disk, stellar halo, satellites) and for their relative contribution to the total amount of EMPs in each MW/M31-like system, by also accounting for the unavoidable galaxy-to-galaxy variations. Following previous theoretical work, we employ a morphological decomposition where both the spatial and kinematic information of the stars are taken into account, which could thus be implemented both for our Galaxy as well as external galaxies equipped with integral-field-spectroscopy observations. 

The TNG50 simulation has been shown to realize well-resolved galaxies that are in many aspects consistent, to reasonable degrees, with observational data. For example, \cite{Pillepich:2019aa} showed that TNG50 returns both star-forming and quiescent galaxies across the mass spectrum, with proportions between the two populations in the ball park of observational findings also at intermediate and high redshift \citep[$z\gtrsim2$;][]{Donnari:2021ab}. \citet{ParkM:2022aa} illustrated that the star-forming galaxies follow well the low-redshift observed main-sequence \citep{Whitaker:2012aa,Renzini:2015aa} in the stellar mass range of $10^{10.5-11.5}\,\msun$ and the TNG50 star-forming main sequence is also consistent with novel observational inferences at $z\sim1$ \citep{NelsonE:2022aa}.
%
%
\citet{Joshi:2021aa} showed that the stellar mass-to-halo mass relation of galaxies with virial masses of $10^{9.5-12}\,\msun$ in TNG50 is consistent with other empirical derivations and observation-based findings \citep{Behroozi:2019aa,Read:2017aa}. 
The radial profiles of the star formation rates in TNG50 galaxies are consistent with those derived from MaNGA in the local Universe \citep{Motwani:2022aa}
and those of 3D-HST galaxies at $z \sim 1$ \citep{NelsonE:2021aa}. Moreover, the gas–phase metallicity gradients within TNG50 star-forming galaxies at $z=0-0.5$ are consistent with observed ones \citep{Hemler:2021aa}.
For MW/M31-like analogues,
\citet{Engler:2021aa} analysed the satellite abundances and found a large scatter across simulated  hosts, but with the median satellite mass function being roughly consistent with the MW or M31. 
\citet{Emami:2022aa} studied the stellar kinematics of MW analogues and demonstrated that the radial profiles of the velocity anisotropy parameter are consistent between TNG50 simulated galaxies and observations.

The theoretical predictions from this paper based on TNG50 therefore represent a useful and statistically-rich benchmark for future observational and numerical studies. We describe the TNG50 simulation and how the MW/M31 analogues are chosen in Secs.~\ref{sec:tng50} and \ref{sec:mwm31}. We define the morphological components of the system in Sec.~\ref{sec:morph} and show the results in Sec.~\ref{sec:emp_result}. We discuss the dependence of our results on the metallicity threshold in Sec.~\ref{sec:emp_dis}. Finally, we conclude this work and summarize our findings in Sec.~\ref{sec:emp_conclu}.


\section{Methods}
\label{sec:emp_met}
In the following sections, we describe the TNG50 simulation and how the MW/M31-like systems are identified. We also explain the post-processing procedure to partition each system into stellar morphological components.

\subsection{The TNG50 simulation}
\label{sec:tng50}
TNG50 \citep{Nelson:2019ab,Pillepich:2019aa} is the highest resolution simulation in the IllustrisTNG\footnote{The simulations of the IllustrisTNG project are fully publicly available and described in \citet{Nelson:2019aa}.} simulation suite \citep{Marinacci:2018aa,Naiman:2018aa,Nelson:2019aa,Pillepich:2018ab,Springel:2018aa}. 
The simulation starts at $z=127$ and runs until $z=0$ following the \citet{Planck15XIII} cosmology ($h = 0.6774$, $\Lambda_{\Omega,0} = 0.6911$,  $\Lambda_{\rm m,0} = 0.3089$, $\Lambda_{\rm b,0} = 0.0486$, $\sigma_8 = 0.8159$, and $n_{\rm s} = 0.9667$). It has a dark matter mass resolution of $3.1 \times 10^5\,h^{-1}\msun$ and a baryonic mass resolution of $5.8 \times 10^4 \,h^{-1}\msun$ with a comoving volume of $35^{3}\,h^{-3}$Mpc$^{3}$. Stellar particles therefore represent $\lesssim 10^5\,\msun$ mono-age stellar populations. 

TNG50 is run with the moving mesh code \textsc{arepo} \citep{Springel:2010aa}. It includes a large number of physical processes such as primordial and metal-line cooling, heating by the extragalactic UV background, stochastic, gas-density threshold-based star formation, evolution of stellar populations represented by star particles, chemical feedback from supernovae and AGB stars, and supermassive black hole (SMBH) formation and feedback. The details of the model are described in \citet{Weinberger:2017aa} and \citet{Pillepich:2018aa}. Importantly, TNG50 is a cosmological simulation, whereby the coupled equations of gravity, magnetohydrodynamics, and galaxy formation processes are solved in an expanding universe, so that also the hierarchical growth of structure is fully accounted for. Moreover, stellar particles are allowed to age, to loose mass and to enrich the surrounding inter-stellar medium (ISM) with metals, so that subsequent episodes of star formation in the simulated galaxies produce a progressively more metal-enriched ISM and star particles with progressively higher initial metallicities.

\subsection{MW/M31 analogues in TNG50}
\label{sec:mwm31}
To identify analogues of the MW and M31 in TNG50, we follow the criteria already proposed by \citet{Engler:2021aa,Pillepich:2021aa} and described in great detail by Pillepich et al. (in preparation):
\begin{enumerate}
    \item the galaxy has a stellar mass, $M_* (< 30$~kpc), in the range of $10^{10.5-11.2}\msun$, 
    \item its 3D minor-to-major axial ratio $s$ of the stellar mass distribution is less than 0.45 or it appears disky by visual inspection. 
    \item there are no other galaxies with $M_* > 10^{10.5}\msun$ at a distance of less than 500\,kpc,
    \item the mass of the host underlying dark matter halo is $< 10^{13}\msun$.
\end{enumerate}
This leads to the identification of 198 MW/M31-like galaxies in TNG50 at $z=0$, including a majority with stellar bars \citep[][Pillepich et al. in preparation]{Frankel:2022aa}, with more or less massive and extended bulges \citep{Gargiulo:2022aa}, and with a diversity of past merger histories \citep{Sotillo-Ramos:2022aa}. 

Each of these galaxies is surrounded by a more-or-less numerous population of satellite galaxies \citep{Engler:2021aa}: throughout this paper, we consider as MW/M31-like satellites those galaxies identified by the \textsc{subfind}
algorithm \citep{Springel:2001aa,Dolag:2009aa} that fulfill the following criteria:
\begin{enumerate}
    \item the galaxy is within 300\,kpc (3D) of one of the MW/M31 analogues,
    \item it has a stellar mass $\ge 5 \times 10^{6}\msun$ within 2 times its stellar half-mass radius. 
\end{enumerate}
We limit the distances of satellites to the host galaxy to the value typically adopted when quantifying the demographics of the MW's and M31's satellites \citep{McConnachie:2012aa, McConnachie:2018aa}.
Moreover, 300 kpc is slightly larger, but similar, to the typical virial radius of MW/M31-mass haloes. 
In TNG50 a minimum stellar mass of $5 \times 10^{6}\msun$ is equivalent to resolving a galaxy with at least 63 star particles: we impose this minimum as it ensures completeness in the underlying halo mass \citep[see Fig.~A2 in ][]{Engler:2021aa}. 
The number of satellites that fulfil the above selection criteria varies among the 198 MW/M31 analogues from 0 to 20 \citep[see Fig.~3 in][]{Engler:2021aa}. 

\subsection{Morphological decomposition with kinematics}
\label{sec:morph}
Many different methods have been used and proposed in the literature, both observationally and numerically, to subdivide the bodies of galaxies into different stellar components, such as disk, bulge, and stellar halo. Even in the context of the IllustrisTNG simulations, a number of complementary approaches have been used and catalogued \citep[e.g.][]{Du:2019aa, Gargiulo:2022aa, Zhu:2022aa}. Here we follow the recommendation by \citet{Du:2019aa} and go beyond photometry-based (i.e. geometric) morphological decompositions of galaxies. In particular, we embrace the method of \citet{Zhu:2022aa}, which accounts for the stars' kinematics and that we describe below.

Firstly, for each simulated MW/M31-like system, we take all of its stars within the halo radius of $r_\mathrm{halo} = 300$\,kpc. In other words, we only take into account stars that are within 300\,kpc of the centre of a MW/M31 analogue, where the centre is taken to be the location of the particular cell with minimum gravitational potential energy. From these, we exclude satellite stars: these are stellar particles that are gravitationally bound to other galaxies, i.e. satellites, according to the \textsc{subfind} algorithm. The remaining stars compose what we refer to as the ``main galaxy body'': their coordinates and velocities can be translated and hence expressed immediately in the coordinate system of the galaxy -- the bulk velocity of the galaxy is the resultant velocity of all its resolution elements in the reference system of the simulation box. 

Therefore, we classify stars in the main galaxy body into four components based on their circularity $\epz$ (defined below) and their distance from the center of the galaxy $r_*$:
\begin{enumerate}
    \item Cold disk: $\epz > 0.7$ and $r_* \leq r_\mathrm{disk}$; \\
    \item Bulge: $\epz \leq 0.7$ and $r_* < r_\mathrm{cut}$; \\
    \item Warm disk: $0.5 < \epz < 0.7$ and $r_\mathrm{cut} \leq r_* < r_\mathrm{disk}$; \\
    \item Stellar halo:  $\epz \leq 0.5$ and $r_\mathrm{cut} \leq r_* \leq r_\mathrm{halo}$ plus $\epz > 0.5$ and $r_\mathrm{disk} < r_* \leq r_\mathrm{halo}$.
\end{enumerate}
Here, $r_\mathrm{cut} = 3.5$\,kpc, $r_\mathrm{disk} = 6 \times r_\mathrm{disk\,scale\,length}$ is the disk radius, where $r_\mathrm{disk\,scale\,length}$ is the disk scale length computed by \citet{Sotillo-Ramos:2022aa}, and $r_\mathrm{halo} = 300$\,kpc is the maximum distance that we consider.
The separation between bulge and stellar halo at the fixed cut of $r_\mathrm{cut} = 3.5\,$kpc is taken from \citet{Zhu:2022aa}, who showed that the distribution of stars with $\epz < 0.5$ peaks at $< 1.5$\,kpc for most galaxies in TNG50 and most of the stars with $\epz < 0.5$ are at $r < 3.5\,$kpc (see their Fig.~3). 
The idea of this kinematically-defined morphological decomposition is shown with a cartoon plot in Fig.~\ref{fig:morph_decomp}, together with additional details related to how much stellar mass is assigned to each component for all 198 MW/M31-like galaxies of TNG50 (see Appendix \ref{app:emp_app}). 

Next, we list the steps of how we obtain the circularity $\epz$ for each stellar particle of a given galaxy:
\begin{enumerate}
    \item Shift the origin to the centre of the main galaxy such that  $\textbf{r}_\mathrm{*} = \textbf{r}_\mathrm{*, ini} - \textbf{r}_\mathrm{main, ini}$ and remove the bulk velocity so that $\textbf{v}_\mathrm{*} = \textbf{v}_\mathrm{*, ini} - \textbf{v}_\mathrm{main, ini}$, where $\textbf{r}_\mathrm{main, ini}$ and $\textbf{v}_\mathrm{main, ini}$ denote the distance vector to the origin and velocity vector of the main galaxy, respectively.
    \item Compute the un-rotated angular momentum of each star in the main galaxy, $\textbf{j}_\mathrm{*} = \textbf{r}_\mathrm{*} \times \textbf{v}_\mathrm{*}$.
    \item Sum up orbital angular momenta of all stars in the main galaxy $\textbf{J}_\mathrm{gal} = \Sigma_{i} \textbf{j}_{\mathrm{*}}$.
    \item Rotate the coordinate system such that the new $z$-axis is parallel to the angular momentum of the main galaxy; i.e.\ $\hat{\mathrm{z}} \parallel \textbf{J}_\mathrm{gal}$. 
    \item Take the $z$-component of the specific stellar angular momentum after coordinate rotation for each star in the system, ${j}_\mathrm{z, rot}$.
    \item Calculate the value of the circular velocity for each star $v_\mathrm{c} = \sqrt{G M(< r_\mathrm{*,rot})/r_\mathrm{*,rot}}$, where $r_\mathrm{*,rot}$ is the stellar distance from the centre of the galaxy after coordinate rotation. 
    \item Compute the magnitude of the angular momentum that the star would have if it were in a circular orbit at the stellar distance from the centre of the galaxy, namely $j_\mathrm{c} = r_\mathrm{*,rot} v_\mathrm{c}$.
    \item Finally, we obtain the circularity by $\epz \equiv \frac{\textbf{j}_\mathrm{z, rot}}{j_\mathrm{c}}$.
\end{enumerate}
Stars on a circular orbit have $\epz\sim1$, while those in hot and chaotic orbits have $\epz\sim 0$: see Fig.~\ref{fig:epzr} and the Appendix of  \citet{Zhu:2022aa}.

The fifth component of each MW/M31-like system is composed by its satellites, which we consider as a population, i.e. for each MW/M31 analogue and when referring to its satellites or satellites component, we sum up the stars in its satellites (see above). We discuss the scatter across the whole population of MW/M31-like satellites (about 1200 in total) and possible trends as a function of their stellar mass in Sec.~\ref{sec:emp_dis}. Finally, we characterise each MW/M31-like system based on five stellar components: cold disk, bulge, warm disk, stellar halo, and satellites. 

\subsection{EMP stars in TNG50, and across morphological components}
As well as tracing the overall metallicity,
the TNG50 simulation traces individual abundances of 9 species in addition to a tracer of Europium \citep{Pillepich:2018aa, Naiman:2018aa}: H, He, C, N, O, Ne, Mg, Si, Fe. We label star particles as EMP stars if their $\feh$ is $< -3$, measured from their total metal mass fraction at birth. We address different choices of such a threshold in Sec.~\ref{sec:emp_dis}. It is important to keep in mind that star particles in TNG50 are of the order of $10^5\,\msun$ and therefore do not represent individual stars but rather star clusters or populations that form at the same time in the same environment. 

In addition to the spatial selection of stars outlined above, here we consider all star particles that survive until $z=0$.
We compute the EMP mass and stellar mass in each galaxy stellar morphological component to study where we are more likely to find EMP stars. In this work we use two different stellar mass fractions to quantify such statement, which we calculate for each 198 TNG50 MW/M31-like system:
\begin{itemize}
\item frequency by component: \empall fraction, where we take the ratio between the mass in EMPs in each component and the total stellar mass of the same component;\\
\item contribution by component: \empempall, which represents the fraction of EMP mass in each component to the total EMP mass in the system, i.e. across all its components.
\end{itemize}
Here $\rm{comp}$ stands for cold disk, bulge, warm disk, stellar halo, and satellites. The first quantity can also be defined for the whole system. 

\section{Results for TNG50 MW/M31-like galaxies}
\label{sec:emp_result}
Equipped with the output of the TNG50 simulation, with a galaxy selection, and with the morphological decomposition described above, here we present the main results of our analysis. We quantify direct outputs such as the spatial distribution and radial profiles of EMP stars as well as derived results such as the EMP fractions in different morphological components. However, before doing so, we briefly characterise and comment on the MDF of the 198 MW/M31-like galaxies of TNG50 at $z=0$.

\begin{figure*}
    \centering
    \includegraphics[width=0.85\textwidth]{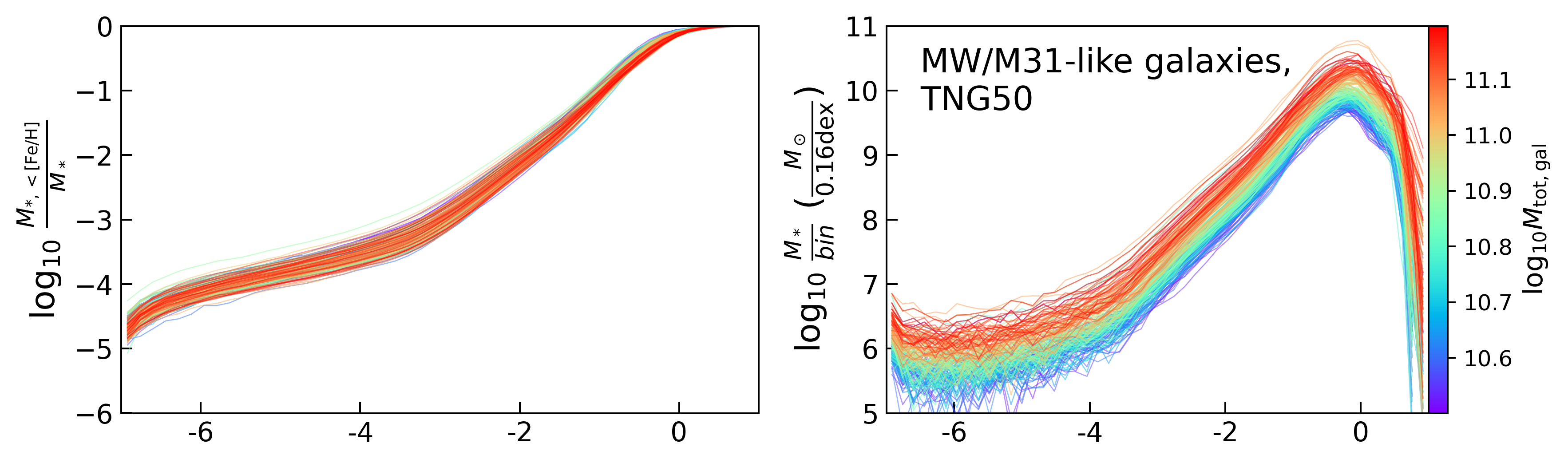}
    \includegraphics[width=0.85\textwidth]{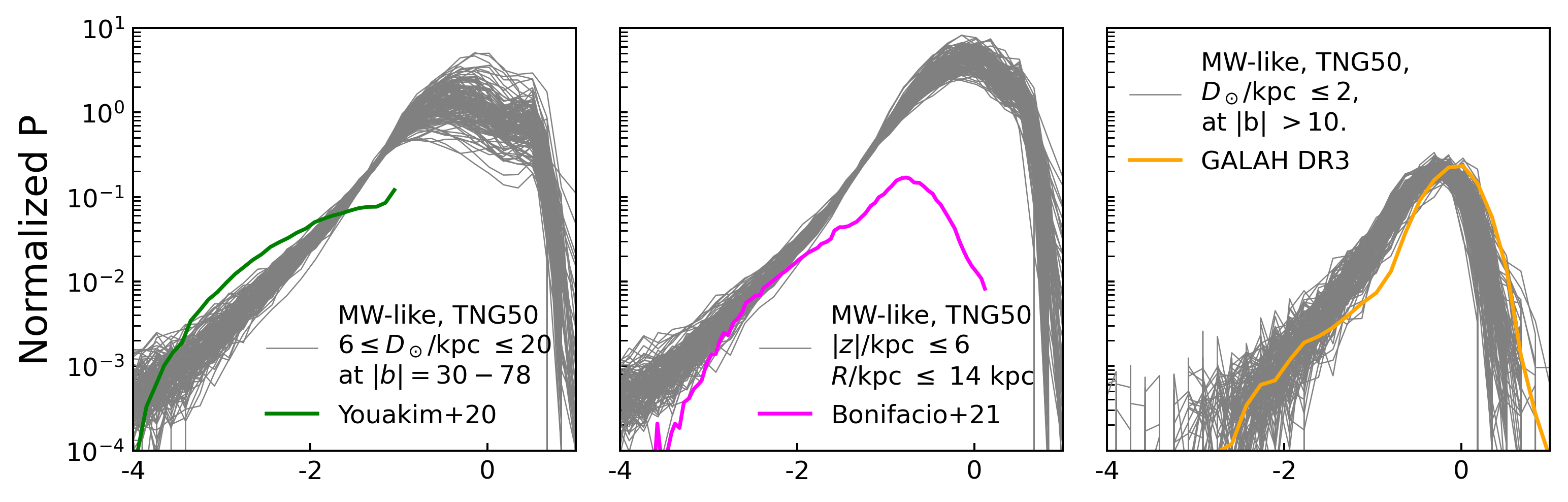}
    \includegraphics[width=0.85\textwidth]{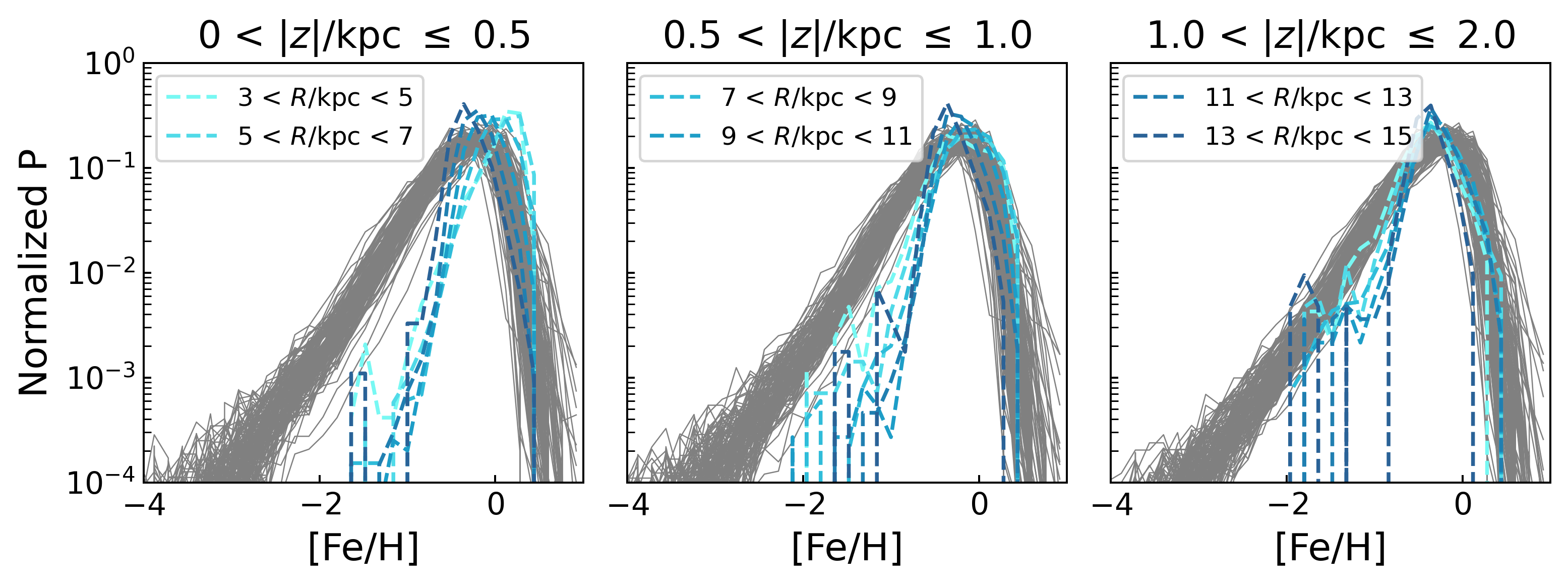}
    \caption[MDFs of TNG50 MW/M31 analogues]{{\bf Stellar metallicity distribution functions (MDFs) of MW/M31-like galaxies in TNG50.} {\it Top panels:} we show the MDFs of all 198 MW/M31 systems, across {\it all} their morphological components: disks, bulge, stellar halo and satellites, colour coded by galaxy stellar mass. The redder the colour, the higher the galaxy stellar mass. On the left, we show the cumulative fraction and on the right, we show the stellar mass. {\it Middle three panels:} MDFs of the subsample TNG50 galaxies with mass more similar to the Milky Way (gray curves) overlaid to results from observations and thus with stars selected by height and radial distance to attempt to account for the surveys' selection functions: we report here the Milky Way's MDF by \citet[][green]{Youakim:2020aa}, \citet[][magenta]{Bonifacio:2021aa}, and by \citet[][orange]{Buder:2021aa}. For the middle left panel, we select stars that have heliocentric distances between 6 and 20 kpc, and at Galactic latitudes between 30 and 78 degrees ($|b| = 30-78$). The probability is then normalised to the total number of stars between $-4. < \feh < -1.05$ to compare with \citet{Youakim:2020aa}. In the middle central panel, we select stars that are < 6\,kpc from the galactic plane and < 14\,kpc from the galactic centre. The probability is again normalised to the total number of stars between $-4. < \feh < -1.05$ to compare with \citet{Bonifacio:2021aa}. For the middle right panel, we select stars that have heliocentric distances $<2$ kpc and that are positioned at Galactic latitudes larger than 10 degrees ($|b| > 10$) to compare with data from GALAH survey Data Release 3 \citep{Buder:2021aa}. In order to take into account the distance-selection criterion from the survey, the solar position in each MW-like galaxy is randomly sampled, at 8.2 kpc from the galactic centre, and we compute the mean MDF over 100 possibilities. The probability is normalised to the total number of stars. {\it Bottom three panels:} MDF comparison to data from \citet[][shades of blue, for three different height selections]{Hayden:2015aa}. We group stars at different distances from the galactic plane to compare with \citet{Hayden:2015aa}, where the probabilities are normalised to total number of stars in each subset. We do not impose any additional selection function to TNG50 star particles but for the aforementioned geometrical spatial cuts that account for the effective footprints of the APOGEE survey Data Release 12. Note that the survey cannot report many metal-poor stars as it hits the detection limit at $\feh \sim -2$.} 
    \label{fig:mdf}
\end{figure*}

\subsection{Metallicity distribution functions of TNG50 MW/M31-like systems}
\label{sec:mdf}

We show the TNG50 predictions for the metallicity distribution functions (MDFs) of all 198 MW/M31-like systems in the top panels of Fig.~\ref{fig:mdf}. Here we consider {\it all} the stars in each galaxy within 300 kpc distance from its centre, without distinguishing by stellar morphological component. On the left, we show the cumulative fraction. On the right, we show the stellar mass in bins of metallicity, without normalising; the colour of the curves denote the galaxy stellar mass. As it can be seen, TNG50 predicts a non-negligible tail towards very low metallicity: $\sim$1 in every 10,000 stellar particles in any given galactic system has a metallicity as low as  $\feh \sim -4$. The galaxy-to-galaxy variation is mostly driven by the different total metal content across galaxies of different mass or assembly history: more massive galaxies have higher MDF peaks and more substantial high metallicity contributions. 

A comparison to observational constraints is not straightforward, as it requires applying to TNG50 data the same selection functions as in the observational surveys, both in terms of survey footprints and possible implicit or explicit selection functions, as in color and magnitude. We partially go in this direction by implementing at least the spatial i.e. geometrical selection functions to TNG50 stars and by comparing to three sets of observational results. This comparison is shown in the lower six smaller panels of Fig.~\ref{fig:mdf}. There we show normalised MDFs for TNG50 MW analogues only (thin grey lines), selected among the 198 galaxies to have stellar mass in the $10^{10.5-11.8}\msun$ range (127 galaxies in total). We select three observational data sets for comparison, for which we adapt the simulated data selection individually.

\begin{itemize}
\item \citet{Youakim:2020aa} aimed to study the metallicity distribution in the Galactic halo. They analysed $\sim 80000$ main sequence turnoff stars from the Pristine Survey that have heliocentric distances between 6 and 20 kpc, and at Galactic latitudes between 30 and 78 degrees ($|b| = 30-78$). Their results are shown as the green solid curve in middle left panel. Their stellar metallicities fall in the range of $\feh = [-4, -1.05]$.

\item \citet[][]{Bonifacio:2021aa}  analysed $\sim 140000$ stars from SDSS data release 12. These stars are located at $\leq 6$ kpc from the Galactic plane and have distances $\leq 14$ kpc from the Galactic center

\item \citet{Buder:2021aa} presented stellar spectra from the GALAH survey Data Release 3 with more than 600,000 stars. The stars are within 2 kpc from the Sun and with Galactic latitudes larger than 10 degrees ($|b| > 10$). We follow the same spatial selection criteria in order to compare the MDFs.

\item \citet{Hayden:2015aa} used APOGEE Data Release 12 to derive MDFs in varying bins of heliocentric distances and heights from the mid Galactic plane ($|z|$/kpc) -- see Sec.~\ref{sec:emp_intro}. The survey did not report many metal-poor stars and hit the detection limit at about $\feh \sim -2$. We show them as dashed curves in the three bottom panels of Fig.~\ref{fig:mdf} and apply similar spatial cuts to the stars of the TNG50 MW-like galaxies.
\end{itemize}

To make a somewhat fairer comparison with the MDFs in \citet{Youakim:2020aa}, \cite{Bonifacio:2021aa} and the GALAH Survey \citep{Buder:2021aa}, we follow the same selection based on spatial information, as described in the caption. However, it should be stressed that there are other selections in the making of the observed MDFs (e.g. in color or magnitude and hence potentially, indirectly, on metallicity) that we are not replicating for the TNG50 stars. Keeping this in mind, we see that the peak of the MDF from \citet{Bonifacio:2021aa} is lower than the one from the TNG50 MW analogues; both MDFs from \citet{Bonifacio:2021aa} and \citet{Youakim:2020aa} show convex curves, whereas the MDF from TNG50 appears to be concave; the MDFs from the GALAH survey and TNG50 MW analogues cover the same metallicity range and agree well in general, while the peak from the GALAH survey is at slightly higher metallicity; and finally, the overall shape of the TNG50 MDFs are wider than the ones in \citet{Hayden:2015aa} at $|z| \leq 1.0$\,kpc, whereby the widths of the MDFs between TNG50 and the \citet{Hayden:2015aa} results are more similar at $1.0 < |z| / \rm{kpc} \leq 2$. 

\begin{figure*}
    \centering
    \includegraphics[width=0.95\textwidth]{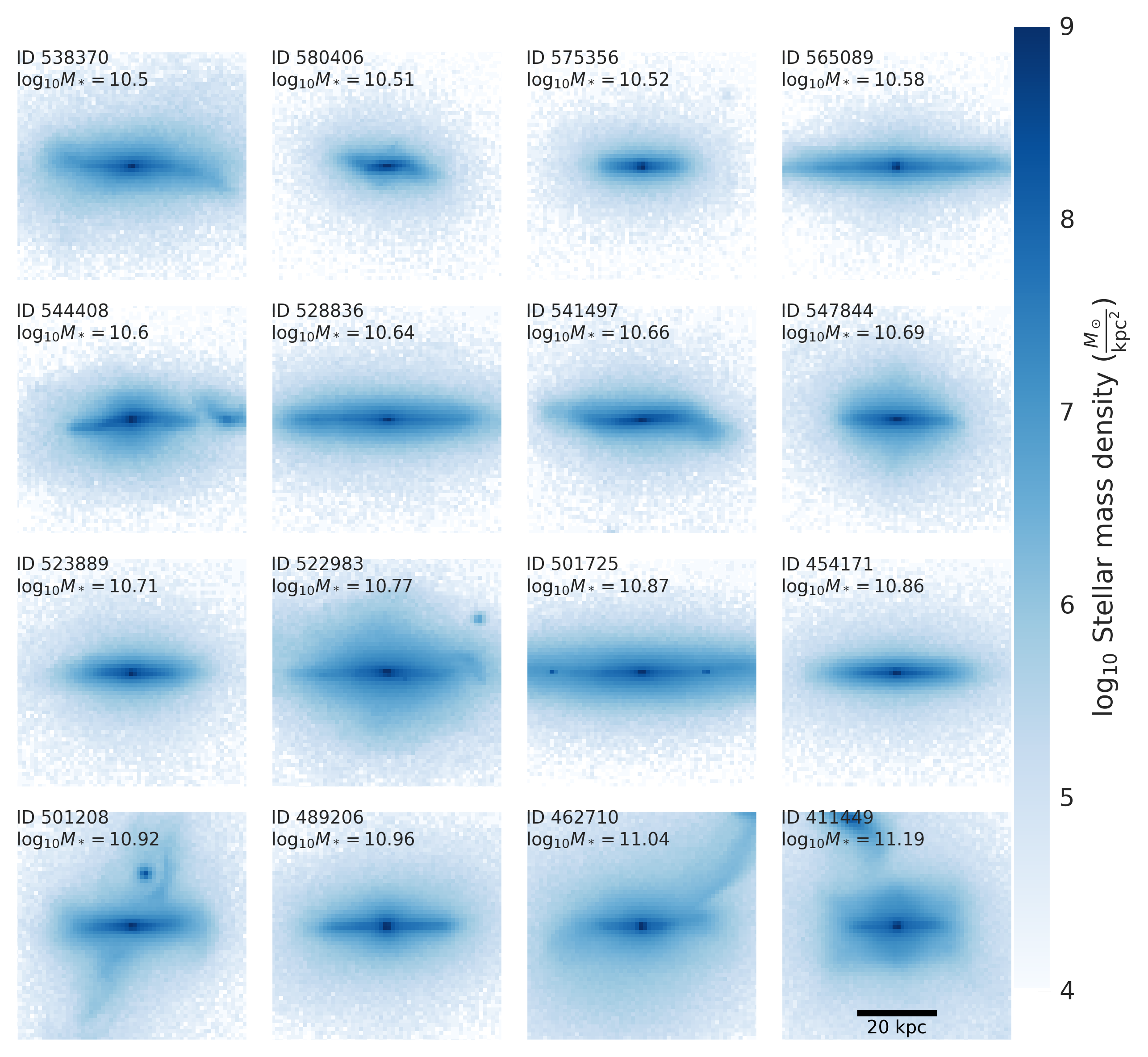}
    \caption[Spatial distribution of stars in 16 MW/M31 analogues]{\textbf{Stellar mass column density of 16 randomly-chosen MW/M31-like galaxies among the 198 of TNG50 at $z=0$.}  The galaxies are shown in edge-on projections, based on the orientation of their stellar disks. The ID is the unique \textsc{subfind} identifier of the galaxy in TNG50 at the snapshot 099. The colours denote the cumulative stellar mass density in spatial pixels of $0.6 \times 0.6$ kpc, for a total of 60 kpc per side. Here we include all stars, irrespective of metallicity.}
    \label{fig:smap}
\end{figure*}
\begin{figure*}
    \centering
    \includegraphics[width=0.95\textwidth]{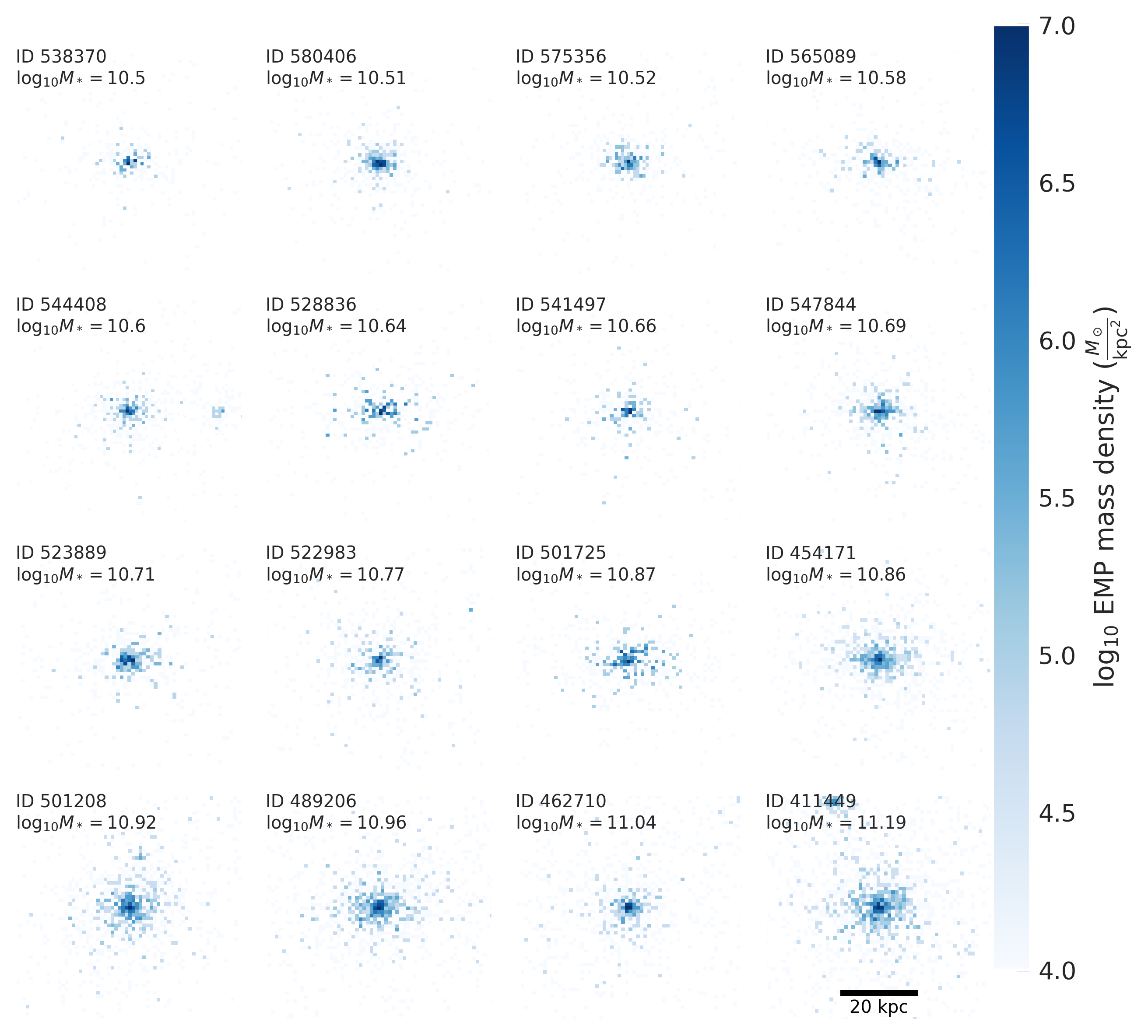}
    \caption[Spatial distribution of EMP stars in 16 MW/M31 analogues]{As in Fig.~\ref{fig:smap} but only for the EMP stellar particles. }
    \label{fig:lzmap}
\end{figure*}

Given the complexity of the observations-to-simulation comparison, the impact of the selection functions on the observationally-derived MDFs (which are indeed all different), and the fact that the MDFs at $z=0$ are the results of 14 billion years of star formation, supernova explosions and stellar winds, outflows triggered by star formation and SMBH feedback, and accretion of stars from lower-mass galaxies and mergers, we conclude that it is reassuring that TNG50 returns MDFs of simulated MW analogues that are in the ball park of the constraints for the Galaxy. We can hence proceed with our analyses with some added confidence in the underlying model. To compare the MDF at lower metallicity (e.g. $\feh \lesssim -3.5$), we would need more complete observational data samples: the distributions of Fig.~\ref{fig:mdf} in the EMP regime are thus predictions of the TNG50 simulation, to be proved or ruled out with future data.

\begin{figure*} 
    \centering
    \includegraphics[width=0.9\textwidth]{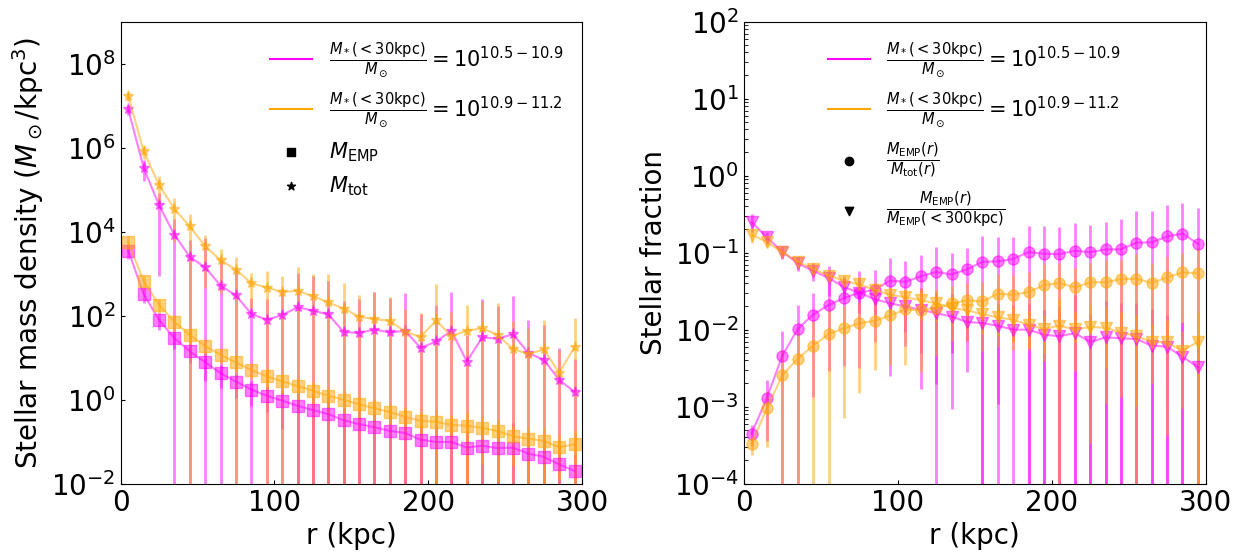}
    \caption[Radial profiles of \empallr fraction and $M_\mathrm{EMP}(r)$-to-$M_\mathrm{EMP, tot}$ fraction]{ \textbf{Radial spatial distribution of EMPs in 198 TNG50 MW/M31 analogues}. Here we consider all stars irrespective of morphological component, including stars in satellites, and we show the unweighted-mean profiles across many simulated galaxies, along with the their galaxy-to-galaxy 1~$\sigma$. In the left panel, we show the radial profiles of the total stellar mass (stars) and of the stellar mass in EMP stellar populations (squares). In the right panel, we show radial profiles of the mean \empallr stellar fraction (circles) and $M_\mathrm{EMP}(r)$-to-$M_\mathrm{EMP}(<300 \rm{kpc})$ stellar fraction (triangles). The radial bins have a width of 10~kpc. The magenta and orange curves denote galaxies in different groups: MW-mass (galaxy stellar mass within 30 kpc of $M_* = 10^{10.5-10.9} \msun$) and M31-mass ($M_* = 10^{11.9-11.2} \msun$), respectively.}
    \label{fig:rdf}
\end{figure*}

\subsection{Spatial distribution of EMP stars}
Most of the EMP stars are located in the central regions of the MW/M31-like systems and do not typically form a visually-identifiable disk. We choose at random 16 examples among the 198 MW/M31 analogues and show the stellar mass column density of all their stars (Fig.~\ref{fig:smap}) and of their EMP stars (Fig.~\ref{fig:lzmap}), both in edge-on projections. In the first set of images, structures like the bulge and the cold disk can be clearly seen with visual inspection.

Next we show the radial distributions of all stars and EMP stars in Fig.~\ref{fig:rdf}. 
We divide the 198 systems into 2 groups: MW-mass where $M_*(<30\mathrm{kpc})/\msun = 10^{10.5-10.9}$ \citep{McMillan:2017aa} and M31-mass where  $M_*(<30\mathrm{kpc}))/\msun = 10^{10.9-11.2}$  \citep{Tamm:2012aa}. These two groups are plotted in magenta and orange, respectively.
In the left panel, we display the mass density of all stars with star symbols and EMP stars with squares at different radii. The mean and 1 standard deviation among the 198 systems are shown.
In the right panel, we show the radial profiles of the \empallr fraction and  $M_\mathrm{EMP}(r)$-to-$M_\mathrm{EMP} (<300\mathrm{kpc})$ at different radii. 
As anticipated in the maps, EMPs are centrally concentrated, with declining mass density profiles similar to those of stars of any metallicity. On the other hand, the \empallr fraction increases as $r$ increases and we observe a clear trend that MW-mass galaxies shows higher \empallr fraction than the M31-mass galaxies at large galactocentric distances. 
Conversely, the fraction of EMP stars located at a certain radius to the total EMP mass ($M_\mathrm{EMP}(r)$-to-$M_\mathrm{EMP} (<300\mathrm{kpc})$) decreases as the galactocentric distance increases and there is no significant difference among galaxies of different masses. 

In other words, EMPs are more frequent and easier to find at large galactocentric distances, with one out of every 10-100 stars being EMP beyond a few tens of kpc in comparison to one EMP star every $10^{3-4}$ stars in the more inner regions. On the other hand, cumulatively, the inner regions of galaxies still contribute relatively more to the total mass in EMPs than the outskirts. The distribution of EMP among different morphological components is discussed in more details in Sec.~\ref{sec:emp_empfreqmorph}.

\subsection {EMP frequency and contribution in and by different components}
\label{sec:emp_empfreqmorph}
In this section we quantify the frequency and contribution of EMPs from the different morphological components of TNG50 MW/M31-like systems. 

\begin{figure*}
    \centering
    \includegraphics[width=0.5\textwidth]{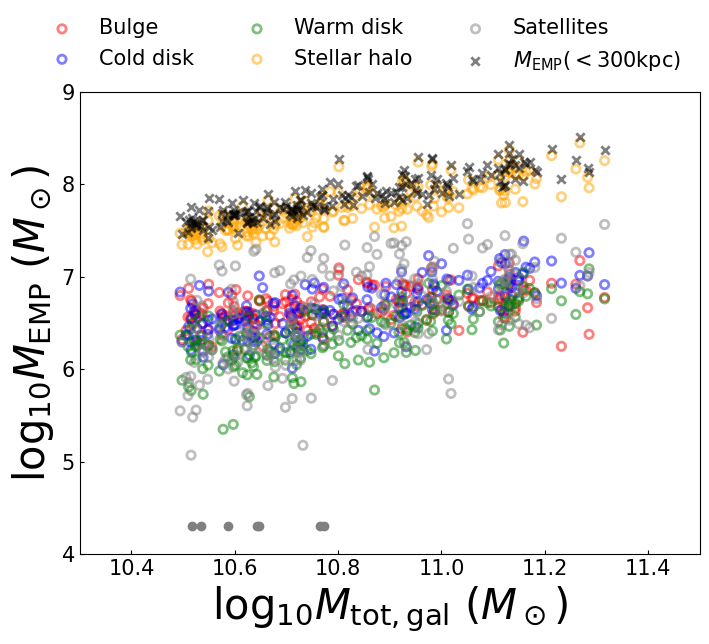}
    \includegraphics[width=0.9\textwidth]{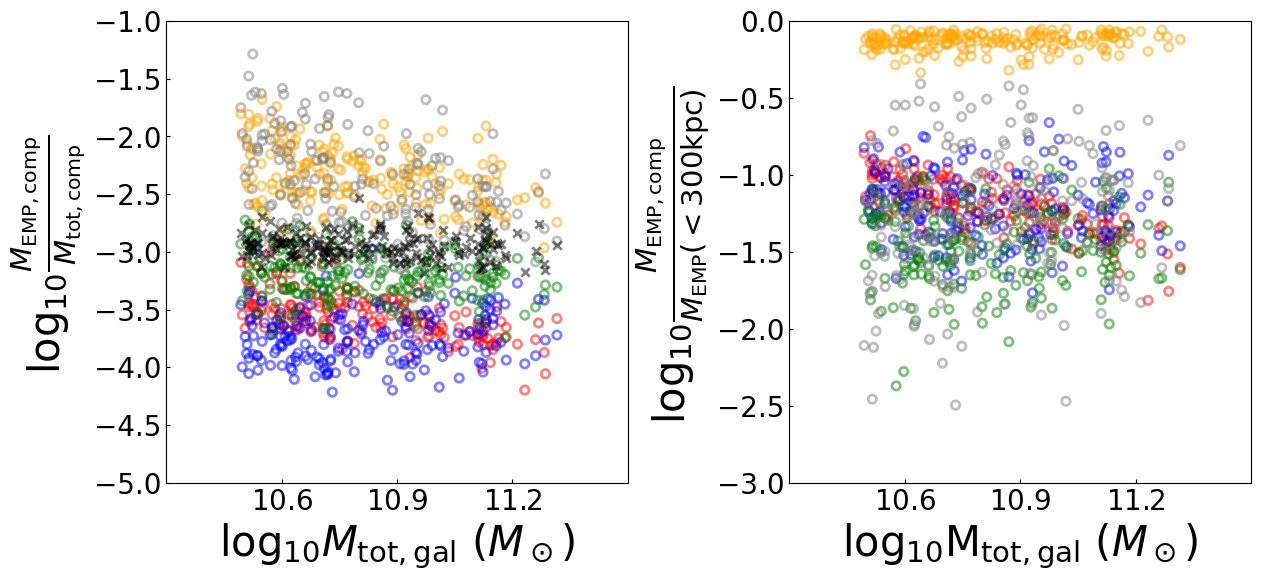}
    \caption[Mass fraction of EMP stars in 198 TNG50 MW/M31 analogues in different morphological components]{\textbf{Stellar mass and mass fractions of EMPs in 198 TNG50 MW/M31 analogues across their different morphological components.} In the top panel, we give the amounts of stellar mass in EMP stellar populations predicted by TNG50 across and within the different galaxies. In the bottom left panel, we show the \empall fraction, i.e. the frequency of EMPs on a component-by-component basis. In the bottom right, we show the \empempall fractions, i.e. the contribution of each morphological component to the total mass in EMPs, i.e. across all components. Satellites belonging to each galaxy are considered as one component in the system (the ``satellites'', see Sec.~\ref{sec:morph} for details). Bulge, cold disk, warm disk, stellar halo, and satellites in each MW/M31-like system are shown as red, blue, green, orange, and grey circles, respectively. The amount of EMPs across all the morphological components ($M_\mathrm{EMP} (<300\mathrm{kpc})$) is shown as black crosses. 
    There are 6 MW/M31 analogues without any satellites and 1 MW/M31 analogue that has only one metal-enriched satellite. They are manually added in the bottom of the top panel in filled grey circles.
    }
    \label{fig:empfraction}
\end{figure*}

In the top panel of Fig.~\ref{fig:empfraction} we show the TNG50 predictions for the mass in EMPs in MW/M31-like systems throughout their bodies and haloes ($M_\mathrm{EMP} (<300\mathrm{kpc})$, black crosses) and within different morphological components (coloured circles). In the lower panels, we
quantify, on the left, the \empall fraction of each component vs the stellar mass of the main galaxy (i.e.\ the EMP frequency within each component) and, on the right, the fraction of EMP mass in each component to the total EMP mass of the system (\empempall), namely the contribution of EMPs by each component to the total EMP mass.

According to TNG50, the stellar halo of the main galaxy hosts the great majority of the EMP stars in all MW/M31-like systems (orange circles in all panels of Fig.~\ref{fig:empfraction}). In fact, the average EMP frequency is similar within the stellar halo and the satellites for most of the systems (bottom left panel, orange vs. grey circles), even though there is typically much less mass in EMPs in the satellite populations than in the stellar haloes. Also keep in mind that the frequency and contribution of EMPs of satellites can vary by up to two orders of magnitude depending on the system. More specifically, in all simulated MW-M31-like objects, the mass in EMPs is about 1/1000 of the total stellar mass (black crosses in the lower left panel), and the frequency of EMPs in stellar halo and satellites is on average one every 100-300 stars, with a mild dependence on galaxy stellar mass. 

On the other hand, EMPs are relatively rarer within the bulges and cold disks of MW/M31-like galaxies (red and blue circles in the lower left panel of Fig.~\ref{fig:empfraction}). Yet, given the large stellar mass of these components, their overall contribution of EMP mass to the total galaxy-wide EMP mass is not negligible. About ten per cent of the total EMP mass in MW-mass galaxies can reside in their bulge, even though this fraction decreases as the stellar mass of the main galaxy increases (red circles in lower right panel). Even more interestingly, there are TNG50 MW/M31-like systems whose cold disks also host non-negligible amounts of EMPs (blue circles in lower right panel):
33 MW/M31-like galaxies in TNG50 have cold disks that contribute more than 10 per cent
to the total EMP mass, each with $\gtrsim 10^{6.5-7}\,\msun$
of EMPs in cold circular orbits. These can provide theoretical counterparts for understanding the origin of observed EMPs with near-circular orbits in the Galaxy (see Sec. \ref{sec:emp_intro}).

\section{Discussion}
\label{sec:emp_dis}

\subsection{EMPs across TNG50 MW/M31-like satellites}
\begin{figure}
    \centering
    \includegraphics[width=\columnwidth]{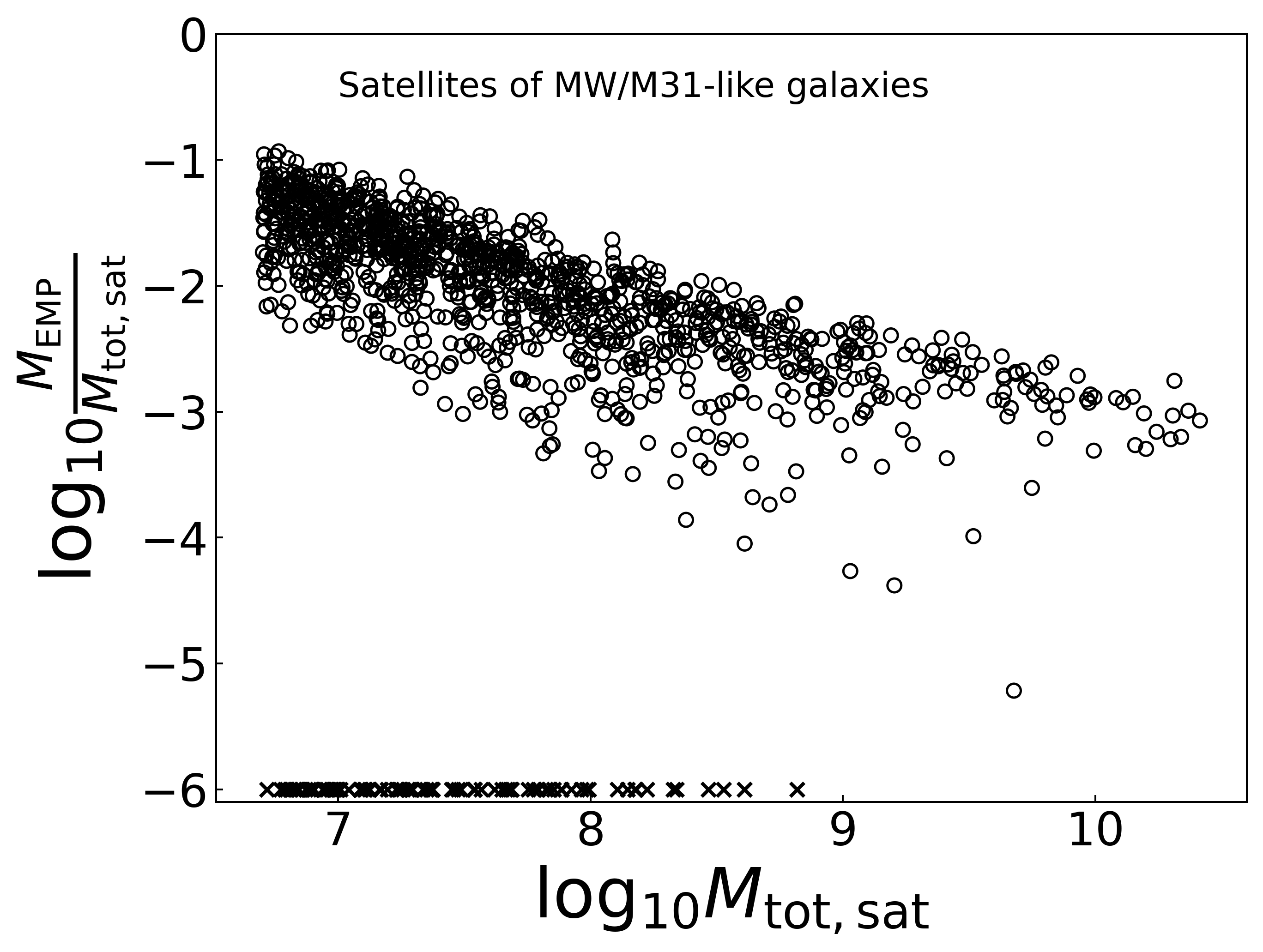}
    \caption[Mass fraction of EMPs in MW/M31-like satellites]{{\bf Stellar mass fraction of EMPs in TNG50 MW/M31-like satellites, across all the 198 TNG50 MW/M31-like hosts.} This is shown as the mass fraction to the total stellar mass of the satellite, as a function of the latter. In this work, satellites are limited to those within 300 kpc from the host centre (3D distance) and to the massive end of the classical dwarfs of the Milky Way. Satellites with no EMPs or with too low amounts in comparison to the minimum mass of the stellar particles in the TNG50 simulation (a few $10^4\, \msun$) are placed by hand at the bottom of the plot. 
    }
    \label{fig:empsat}
\end{figure}

Throughout this analysis, we have considered all satellite stars in each MW/M31-like system as one component: namely, we have stacked all satellites together in each system (see Sec.~\ref{sec:morph}). However, the number of EMPs in satellite galaxies is also expected to depend on the properties of each satellite. In Fig.~\ref{fig:empsat} we therefore show the $M_\mathrm{EMP}$-to-$M_{\rm tot, sat}$ fraction for individual satellites. 

According to TNG50, there is a clear trend whereby the fraction of EMPs mass decreases as the stellar mass of the satellite increases. However, based on the findings of Fig.~\ref{fig:empfraction} (bottom left panel, black crosses), such a trend is not observed at even higher galaxy masses, as across the MW/M31-like host sample the frequency of EMPs does not depend on galaxy mass. The scatter also seems to increase as we look at more massive satellites. However, this is an effect of numerical resolution, as there are satellites with a too low amount of EMPs in comparison to the minimum mass of the stellar particles in the TNG50 simulation (a few $10^4\, \msun$) -- these are placed by hand at the bottom of the plot.

\begin{figure}
    \includegraphics[width=\columnwidth]{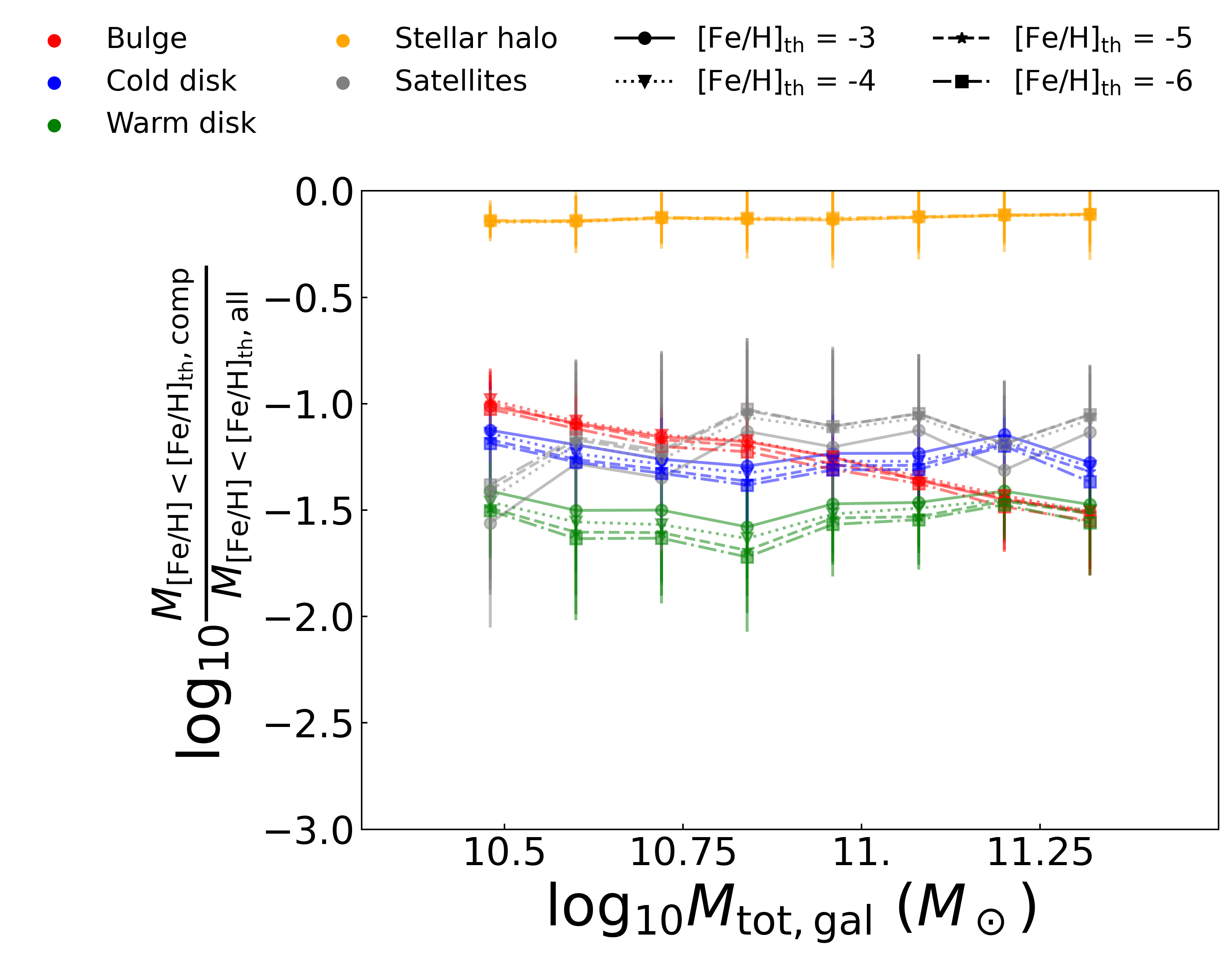}
    \includegraphics[width=\columnwidth]{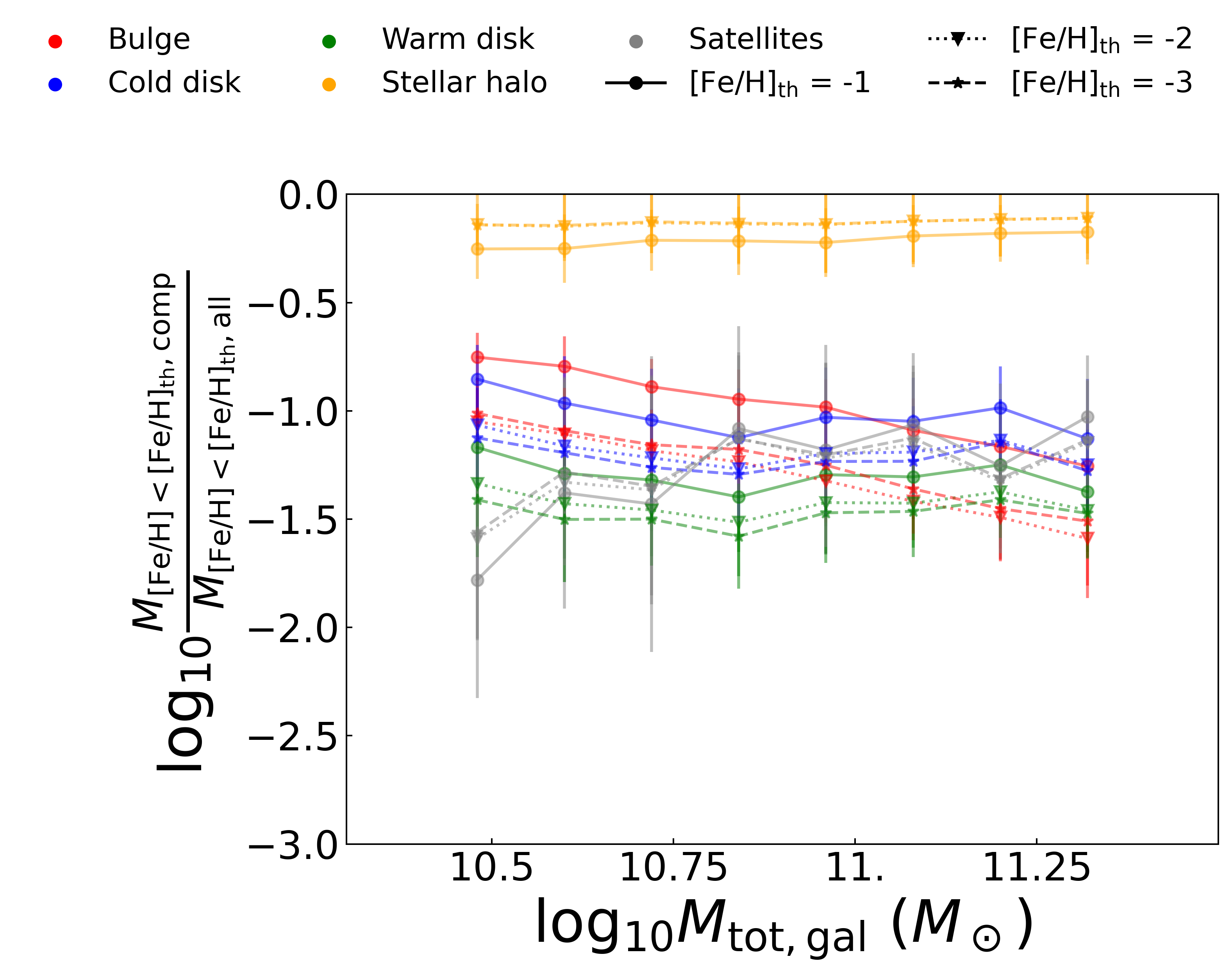}
    \caption[Location of EMPs for alternative definitions of EMPs.]{ \textbf{Location of EMPs for alternative definitions of EMPs in TNG50 198 MW/M31-like systems.} \textit{Top}: \empempall fraction in different morphological components for $\feh_\mathrm{th} = [-3, -4, -5, -6]$ with solid lines, dotted lines, dashed lines, and dot-dashed lines, respectively. The components are coloured as in Fig.~\ref{fig:empfraction}. \textit{Bottom}: fraction in different morphological components with $\feh_\mathrm{th}$ for $\feh_\mathrm{th} = [-1, -2, -3$] with solid lines, dotted lines, and dashed lines, respectively.}
    \label{fig:zthcomp}
\end{figure}
\subsection{Dependence on the definition of EMPs} 
Next, we check whether the \empempall fraction changes with different metallicity thresholds $\feh_\mathrm{th}$ adopted to define extremely metal poor stars. 

In addition to the fiducial value of $\feh_\mathrm{th} = -3$, we plot three cases where $\feh_\mathrm{th} = -4, -5, $ and $-6$ in Fig.~\ref{fig:zthcomp}, top panel. 
The stellar halo component still hosts the great majority of EMPs and the difference among the four cases is negligible.
We find that the \empempall fraction decreases in the the bulge, cold disk and warm disk components as we lower the threshold, whereas the \empempall fraction in the satellites increases. 

For comparison, we also perform the same analysis but increasing the threshold, towards less extreme definitions of metallicity threshold. In Fig.~\ref{fig:zthcomp}, bottom panel, we show \empempall fraction for $\feh_\mathrm{th} = -1, -2, $ and $-3$. We notice that the contribution of stellar halo EMPs to the total amount of EMPs drops from $\sim 0.7$ to $\sim 0.6$ when $\feh_\mathrm{th} = -1$, but still the stellar haloes remain the most most dominant contributors of metal-poor stars in MW/M31-like systems. 

\subsection{Possible limitations of the TNG50 expectations}
Despite being the highest resolution simulation among the suite, TNG50 still cannot resolve mini-haloes ($M_\mathrm{halo} = 10^5-10^6\msun$) properly. Therefore, some star formation in mini-haloes, especially at high redshift, may be unresolved and ignored or delayed in the simulation. This probably leads to an underestimate of EMP stars by some fraction, which may be more relevant for the satellites, especially for the ultra-faint dwarf galaxies (UFDs, $M_\mathrm{tot} < 10^5\msun$). 
For satellites defined in this analysis ($M_\mathrm{tot} > 5\times10^{6}\msun$), this is not an issue.

\section{Conclusion}
\label{sec:emp_conclu}
In this work, we have analysed the location at $z=0$ of extremely metal-poor stars (EMPs, defined here as stars with $\feh < -3$) in 198 MW/M31 analogues in the TNG50 cosmological magnetohydrodynamical simulation of galaxies. 

We have decomposed the galactic systems into five kinematically-defined components, namely bulge, cold disk, warm disk, and stellar halo of the main galaxy -- based on the stellar circularity and radius of each stellar particles -- in addition to the satellite population. For the latter, we have summed up the stellar masses of all satellites in each MW/M31 analogue and considered only satellites within 300 kpc from center of the system and with stellar masses larger than $5 \times 10^6 \msun$ at $z=0$. Furthermore, we have defined two important quantities: the \empall (EMP mass to total stellar mass in one component) and the \empempall (EMP mass in one component to total EMP mass in the system) fractions, where ``comp'' stands for the various morphological components and $M_\mathrm{EMP} (<300\mathrm{kpc})$ denotes the total amount of EMPs throughout a whole galaxy. The former fraction tells us, when we look at a certain component, how many of its stars are expected to be EMP, so the EMPs frequency by component. The latter tells us, when given a total EMP mass, in which component we may find the most EMPs (i.e. the contribution of EMPs by galactic component).

In addition to returning a diversity of galaxies whose properties are overall aligned with observational findings (see Sec.~\ref{sec:emp_intro}), TNG50 also returns MW/M31-like galaxies whose metallicity distribution functions are in the ball park of the observed MDF of the Milky Way, as derived by three groups:  \citet{Hayden:2015aa}, \citet{Youakim:2020aa}, and \citet{Bonifacio:2021aa}. Overall, TNG50 is able to reproduce the observed MW's MDF (Fig.~\ref{fig:mdf}), at least at  metallicities higher than  $\feh > -4$ or $\feh \gtrsim -2$, where data is available. This consideration provides additional confidence in the underlying numerical model and thus in the bounty of the theoretical expectations extracted in this analysis. 

Overall, and prior to any morphological decomposition, we have shown that TNG50 predicts that EMPs preferentially occupy the central regions of MW/M31-like systems. Namely, the mass in EMPs, likewise the mass in stars of any metallicity, is centrally concentrated, with 3D-averaged stellar mass density profiles declining with galactocentric distance (Figs.~\ref{fig:lzmap} and ~\ref{fig:rdf}, left panel). On the other hand, the 
\empallr ratio increases with galactocentric distance, whereas the \empempall ratio in a system decreases as the distance increases (Fig.~\ref{fig:rdf}, right panel). 
We further divide the 198 analogues into two groups: MW-mass ($M_*(<30\mathrm{kpc})/\msun = 10^{10.5-10.9}$) and M31-mass ($M_*(<30\mathrm{kpc})/\msun = 10^{10.9-11.2}$). 
Although there is large galaxy-to-galaxy variation in the radial profiles of EMPs mass and mass fractions, MW-mass and M31-mass galaxies show similar average behaviours: 
the most noticeable difference between the two sets of galaxies lies in the \empallr ratio profile, whereby the local fraction of EMPs to the total stellar mass is higher for MW-mass galaxies beyond a few tens of kpc from the centre.

With Fig.~\ref{fig:empfraction} we have shown that the TNG50 stellar halo and satellites components present the highest \empall fraction, i.e. the highest frequency of EMPs: as expected, it should therefore be easier to find EMPs in the stellar haloes and satellites of MW/M31-like galaxies than in their innermost regions or in the disks. According to TNG50, there should be one EMP star for every 100-300 stars in the stellar halo and satellites of MW/M31-like galaxies. Moreover, the stellar haloes host the largest \empall fraction, i.e. the largest contribution of EMPs to the total EMP mass in a galaxy, for all 198 MW/M31-like systems, with an EMPs mass in stellar haloes (Fig.~\ref{fig:empfraction}, orange circles) that is $\sim 1$ dex higher than in all the other galactic components, on average. 

We find a large galaxy-to-galaxy variation for the satellites populations: the frequency and contribution of EMPs of satellites can vary by up to two orders of magnitude depending on the system (Fig.~\ref{fig:empfraction}, grey circles). This is because the abundance of satellites, i.e. the satellite mass function, around MW/M31-like galaxies can be very diverse \citep[][varying from zero to $\sim20$ satellites around each of the 198 TNG50 MW/M31-like galaxies]{Engler:2021aa}, and because the EMPs mass fraction in individual satellites is smaller the higher the satellite's stellar mass (Fig.~\ref{fig:empsat}).

Finally, even though it is not easy to visually identify bulge or disky structures in EMP maps (Fig.~\ref{fig:lzmap}), the shear amount of EMPs in the inner regions of galaxies is non-negligible. Up to $\sim10^7\msun$ of EMPs can be hosted in the bulges of MW-mass galaxies, even though this bulge EMP contribution decreases as the stellar mass
of the main galaxy increases (red circles in the top and lower right panels of Fig.~\ref{fig:empfraction}). Even
more interestingly, there are large galaxy-to-galaxy variations as to how many EMPs can actually be located within the cold disk component (blue circles in the top and lower right panels of Fig.~\ref{fig:empfraction}). A few tens among the TNG50 MW/M31-like galaxies actually have more than $\sim10^{6.5-7}\msun$ in EMPs in cold circular orbits. This could provide hints to better understand the origin of EMPs with near-circular orbits that have been recently observed in the Galaxy. 

The results of this work, which are qualitatively and, to a large extent, also quantitatively independent of the precise definition of extremely metal-poor stars (Fig.~\ref{fig:zthcomp}), provide theoretical predictions for future search of EMPs  and for stars with even lower metallicity: the chance of finding EMPs is the highest in the stellar halo. 

\section*{Acknowledgements}
We thank the anonymous referee for insightful feedback, Alina Böcker, Christoph Engler, and Mattis Magg for useful discussions and inputs.
Special thanks go to Michael Hayden for providing the raw data that allowed us to reproduce the MDFs from \citet{Hayden:2015aa} and Sven Buder for providing the analysis routine and raw data from the GALAH survey Data Release 3 to compare the MDFs.
The authors acknowledge financial support from DFG via the Collaborative Research Center (SFB 881, Project-ID 138713538) 'The Milky Way System' (subprojects A1, B1, B2, B8). 
SCOG and RSK also acknowledge funding from the Heidelberg cluster of excellence (EXC 2181 - 390900948) `STRUCTURES: A unifying approach to emergent phenomena in the physical world, mathematics, and complex data', and from the European Research Council in the ERC synergy grant `ECOGAL – Understanding our Galactic ecosystem: From the disk of the Milky Way to the formation sites of stars and planets' (project ID 855130).
The analysis was carried out on the ISAAC and VERA computers at the Max Planck Computing and Data Facility (MPCDF).
We also acknowledge the HPC resources and data storage service SDS@hd supported by the Ministry of Science, Research and the Arts Baden-W\"{u}rttemberg (MWK) and the German Research Foundation (DFG) through grant INST 35/1314-1 FUGG and INST 35/1503-1 FUGG.

\section*{Software}
\href{https://github.com/illustristng/illustris_python}{illustris\_python}, matplotlib \citep{Hunter:2007aa}, numpy \citep{Harris:2020aa}, pandas \citep{McKinney:2010aa,Reback:2022aa}, python \citep{python09}, scipy \citep{Virtanen:2020aa}.

\section*{DATA AVAILABILITY}
As of February 1st, 2021, data of the TNG50 simulation series are publicly available from the IllustrisTNG repository: \href{https://www.tng-project.org}{https://www.tng-project.org} and described by \citealt{Nelson:2019aa}. Data directly related to content and figures of this publication will be shared upon reasonable request to the corresponding author.




\bibliographystyle{mnras}
\bibliography{Reference.bib}



\appendix
\section{Kinematically-defined morphological  decomposition}
\label{app:emp_app}
\begin{figure}
    \centering
    \includegraphics[width=0.9\columnwidth]{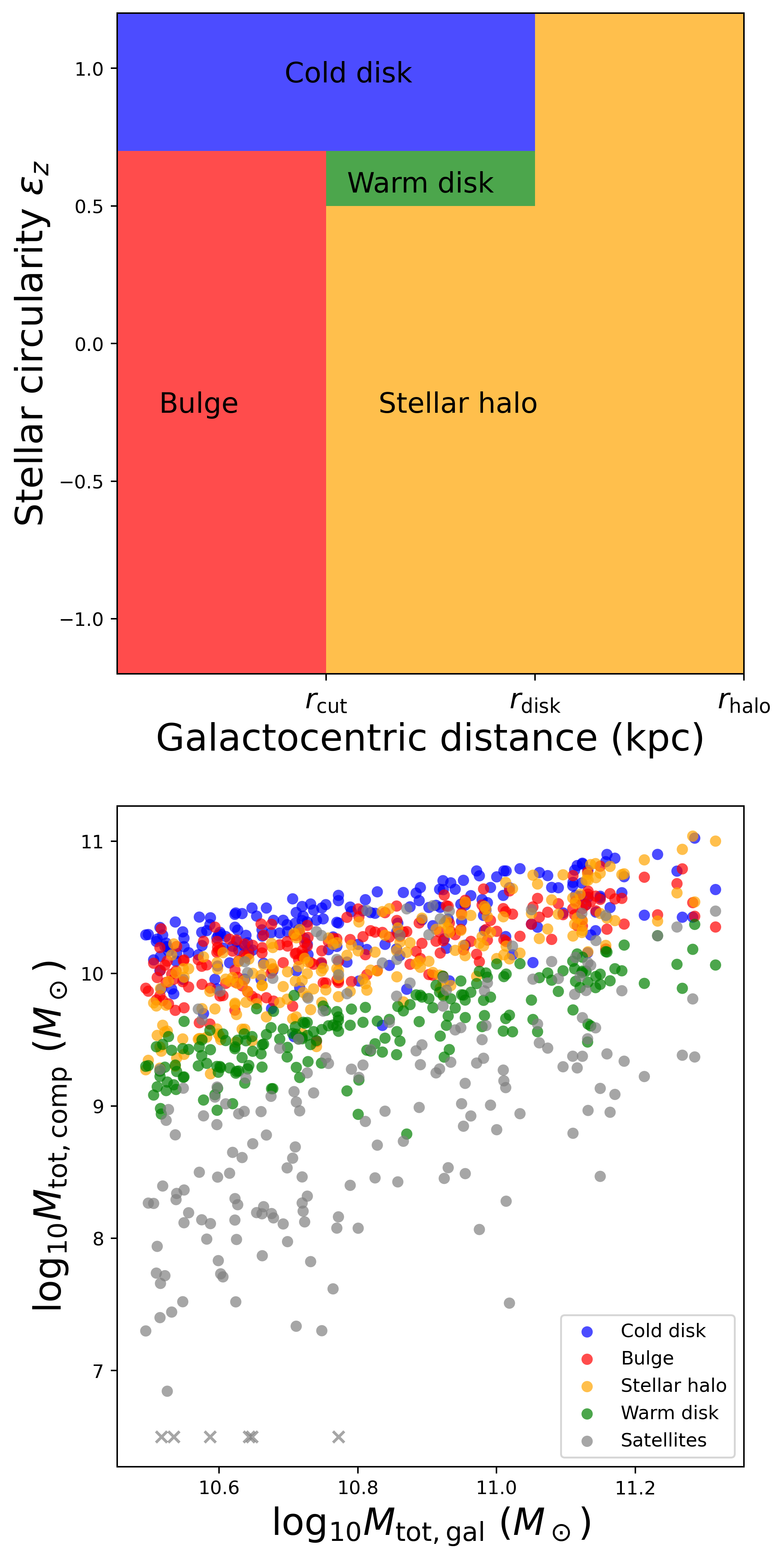}
    \caption[Kinematically-defined morphological decomposition of the simulated MW/M31-like galaxies adopted throughout.] {{\bf Kinematically-defined morphological decomposition of the simulated MW/M31-like galaxies adopted throughout.} {\it Top}: Cartoon plot of the morphological decomposition of the main galaxy's body, based on circularity and galactocentric distance of the stars: $r_\mathrm{cut} = 3.5$\,kpc, $r_\mathrm{disk} = 6 \times r_\mathrm{disk\,scale\,length}$, and $r_\mathrm{halo} = 300$\,kpc. In addition to the depicted components, namely bulge, cold disk, warm disk, and stellar halo, satellite galaxies are identified as gravitationally-bound substructures of stars within 300 kpc from the centre, via the \textsc{Subfind} algorithm. In this paper, we only consider satellites more massive than $5\times10^6 \msun$ in stars.
    {\it Bottom}: stellar mass in the distinct morphological components of all 198 MW/M31-like systems of TNG50 at $z=0$.}
      \label{fig:morph_decomp}
\end{figure}

We expand here on the definition of the kinematically-defined morphological components that we have used throughout the analysis to partition the bodies of the simulated galaxies.
In Fig.~\ref{fig:morph_decomp}, top, we show a cartoon schematic of the decomposition based on the stellar circularity and galactocentric distance of each stellar particle, as described in Section~\ref{sec:morph}. This pertains to the stars that belong to the main galaxy's body, i.e.\ bulge, cold disk, warm disk, and stellar halo, whose stars are all gravitationally bound to the main galaxy. To these, stars in satellites are added and grouped in the satellite component, as described in the main text.
In Fig.~\ref{fig:morph_decomp}, bottom, we give an overview of how much stellar mass is contained in each morphological component, for each 198 TNG50 MW/M31-like galaxy. Among the five components, the cold disk generally has the most mass. The stellar mass in satellites exhibits a much larger scatter: in a few some cases, the stellar mass in satellites is comparable to that of the bulge or cold disk; six systems have no satellites at all -- they are placed by hand at the bottom of the plot using grey crosses. 

\begin{figure*}
    \centering
    \includegraphics[width=1.0\textwidth]{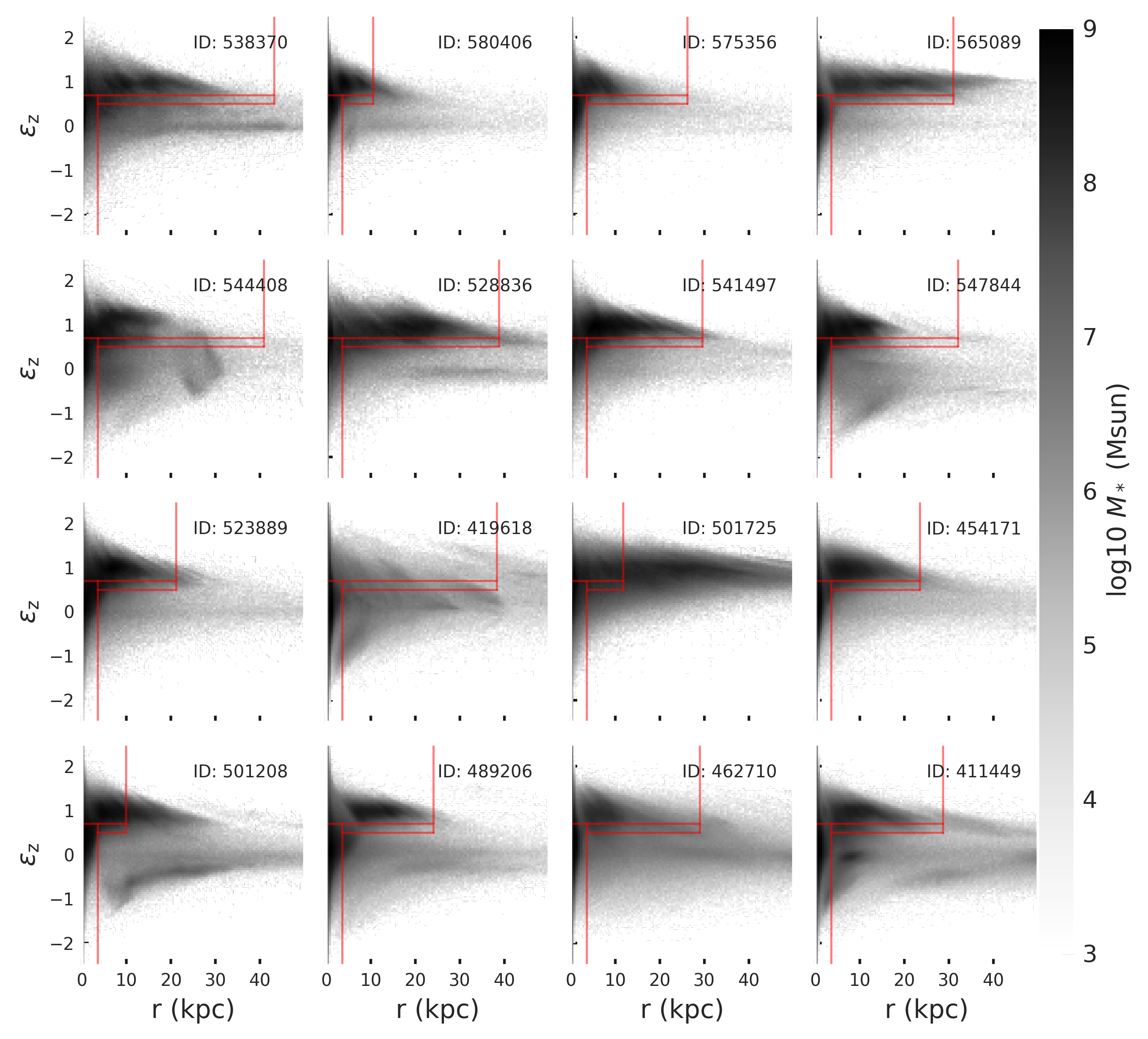}
    \caption[Stellar circularity $\epz$ v.s the distance from the centre of the galaxy in 16 MW/M31 analogues]{ \textbf{Stellar mass density distribution in the stellar circularity $\epz$ v.s. distance from the centre of the galaxy in 16 example TNG50 MW/M31 analogues.} The colorbar shows the cumulative stellar mass along the line of sight in bins of 0.05 in circularity and 0.05 kpc in distance.
    The red lines show the boundaries of the different components as in Fig.~\ref{fig:morph_decomp}.}
    \label{fig:epzr}
\end{figure*}

The cartoon-like schematic of  Fig.~\ref{fig:morph_decomp} is realised in the various systems as shown in Fig~\ref{fig:epzr}, where we plot the stellar circularity $\epz$-radius phase diagram for 16 MW/M31 galaxies as examples. Satellite stars are excluded in this plot. We overplot the boundaries of different components as in Fig.~\ref{fig:morph_decomp} in red.
For the stellar disks of the galaxies visually inspected in Fig.~\ref{fig:smap}, an overdensity around $\epz = 1$ is clearly present, whereas the bulges appear as overdensities around $\epz = 0$ and at small radii, as expected.


\bsp	
\label{lastpage}
\end{document}